%% file: paper.tex
\documentstyle[twoside,fleqn,espcrc2,epsf]{article}
\newcommand\figcaption[1]{\vskip-0.38truein\caption{#1}\vskip0.1truein}

\newcommand\lili{\hbox{$\{ U_i U_i \}$}}
\newcommand\lilj{\hbox{$\{ U_i U_j \}$}}
\newcommand\sli{\hbox{$ \{ S U_i   \}$}}
\newcommand\ssli{\hbox{$ \{ SS, S U_i   \}$}}

\newcommand\MSbar{\hbox{$\overline{MS}$}}
\newcommand\mbar{\hbox{$\overline{m}$}}
\newcommand\GeV{\mathord{\rm \;GeV}}
\newcommand\MeV{\mathord{\rm \;MeV}}

\newcommand\etal{{\it et al.}}

\begin{document}

\title{Status Report on Weak Matrix Element Calculations\thanks{Based on talks presented 
by Rajan Gupta and Tanmoy Bhattacharya.  These calculations have been done on
the CM5 at LANL as part of the DOE HPCC Grand Challenge program, and
at NCSA under a Metacenter allocation.}}

\author{Rajan Gupta and Tanmoy Bhattacharya 
        \address{T-8 Group, MS B285, Los Alamos National
        Laboratory, Los Alamos, New Mexico 87545 U.~S.~A.~}
}

\begin{abstract}
This talk presents results of weak matrix elements calculated from
simulations done on 170 $32^3 \times 64$ lattices at $\beta = 6.0$
using quenched Wilson fermions.  We discuss the extraction of
pseudoscalar decay constants $f_\pi$, $f_K$, $f_D$, and $f_{D_s}$, the
form-factors for the rare decay $B \to K^* \gamma$, and the matrix
elements of the 4-fermion operators relevant to $B_K$, $B_7$,
$B_8$. We present an analysis of the various sources of systematic
errors, and show that these are now much larger than the statistical
errors for each of these observables. Our main results are $f_D=186(29) \MeV$, 
$f_{D_s}=224(16) MeV$, $T_1=T_2=0.24(1)$, $B_K(NDR, 2\GeV)=0.67(9)$, and 
$B_8(NDR, 2\GeV)=0.81(1)$.

\end{abstract}

\maketitle

\makeatletter 

\setlength{\leftmargini}{\parindent}
\def\@listi{\leftmargin\leftmargini
            \topsep 0\p@ plus2\p@ minus2\p@\parsep 0\p@ plus\p@ minus\p@
            \itemsep \parsep}
\long\def\@maketablecaption#1#2{#1. #2\par}

\advance \parskip by 0pt plus 1pt minus 0pt

\makeatother

\section{TECHNICAL DETAILS}
\label{s_tech}

We briefly state the details common to all three quantities discussed
in this talk and refer the reader to \cite{HM95,DC95,Mfflat95} 
for details. Preliminary results based on 100 lattices were
presented at LATTICE94 \cite{POTlat94} and the final analysis will be
presented elsewhere \cite{DC95,MFFfinal}.  

We calculate wall and Wuppertal source quark propagators at five
values of quark mass given by $\kappa = 0.135$ ($C$), $0.153$ ($S$),
$0.155$ ($U_1$), $0.1558$ ($U_2$), and $0.1563$ ($U_3$).  These quarks
correspond to pseudoscalar mesons of mass $2835$, $983$, $690$, $545$
and $431$ $\MeV$ respectively where we have used $1/a=2.33\GeV$ for
the lattice scale.  We construct three types of correlation functions,
Wuppertal smeared-local ($\Gamma_{SL}$) and smeared-smeared
($\Gamma_{SS}$), and wall smeared-local ($\Gamma_{WL}$).  The three
$U_i$ quarks allow us to extrapolate the data to the physical isospin
symmetric light quark mass $\mbar=(m_u+m_d)/2$, while the physical
charm mass is taken to be $C$.  The physical value of strange quark
lies between $S $ and $U_1$ and we use these two points to interpolate
to it.  For brevity we will denote the six combinations of light
quarks $U_1 U_1, \ U_1 U_2, \ U_1 U_3,\ U_2 U_2,\ U_2 U_3,\ U_3 U_3$
by \lilj\ and the three degenerate cases by \lili. The $\Gamma_{SL}$, 
$\Gamma_{SS}$, and 3-point functions have been evaluated at the 5 lowest 
lattice momenta, $i.e.$ $p=(0,0,0), (1,0,0), (1,1,0), (1,1,1), (2,0,0)$.

\noindent{\bf Renormalization Constants}: We use the Lepage-Mackenzie 
tadpole subtraction prescription \cite{Lepage}.  Its implementation
consists of three parts in addition to writing the perturbative
expansions in terms of the improved coupling $\alpha_v$. One, the
renormalization of the quark field $\sqrt Z_\psi$ changes from
$\sqrt{2 \kappa} \to \sqrt{8\kappa_c} \sqrt{1- 3\kappa / 4\kappa_c}$;
second, the perturbative expression for $8\kappa_c$ in $Z_\psi$ is
combined with the coefficient of $\alpha_v$ in the one loop matching
relations to remove the tadpole contribution, and finally the typical
momentum scale once the tadpole diagrams are removed is taken to be
$q^* = 1 / a$, $i.e.$ both the scale at which $\alpha_v$ is evaluated
and the scale at which lattice and continuum theories are matched is
set to $q^* = 1 / a$. We label this scheme $TAD1$ for brevity.  The
difference in results for $q^* = 1 / a$ and $q^* = \pi / a$ is used as
an estimate of systematic errors due to fixing $q^*$. Our data gives 
$\kappa_c =0.157131(9)$ \cite{HM95}.\looseness-1

\noindent{\bf Setting the quark masses}: 
In \cite{HM95} we show that a non-perturbative estimate of
quark mass $m_{np}$, calculated using the Ward identity, is linearly
related to $(1/2\kappa - 1/2\kappa_c)$ for light quarks, so either
definition of the quark mass can be used for the extrapolation.  We
choose to use $m_{np}$, and fix $\mbar$, $m_s$ and $m_c$ as follows.
To get $\mbar$ we extrapolate the ratio $M_\pi^2/M_\rho^2$ to its
physical value $0.03182$. We determine $m_s$ by extrapolating
$M_{\phi} / M_\rho$ to $\mbar$
and then interpolating in the strange quark to match the physical
value. We find a $\sim 20\%$ difference between using 
$M_K^2/M_\pi^2$ or $M_\phi / M_\rho$ to fix $m_s$, which we use as an
estimate of the systematic error.  For $m_c$ we use $\kappa=0.135$ as
we have simulated only one heavy mass. With this choice
the experimental values of $M_D,\ M_{D^*}$ and $M_{D_s}$ lie in
between the static mass $M_1$ (measured from the rate of exponential
fall-off of the 2-point function) and the kinetic mass defined as $M_2
\equiv (\partial^2 E / \partial p^2 |_{p=0})^{-1}$.  The difference 
in final quantities between using $M_1$ and $M_2$ is taken to be an
estimate of the systematic error in fixing $m_c$.

\noindent{\bf The lattice scale \protect\boldmath $a$}: 
To convert lattice results to physical units we use $a(M_\rho)$. As
discussed in \cite{HM95}, the $M_\rho$ data show a small but
statistically significant negative curvature. We get $1/a = 2.330(41)
\GeV$ from a linear fit to \lilj\ points, $2.365(48) \GeV$ including a
$m^{3/2}$ correction term in the fit to the 10 $U_iU_j, SU_i, SS$ points, and
$2.344(42) \GeV$ including a $m^{2}$ term. Since all three estimates
are consistent and the form of the chiral correction cannot be
resolved we use the result from the linear extrapolation and assign
$3\%$ as an estimate of the systematic error.\looseness-1

\section{DECAY CONSTANTS}
\label{s_decay}

The pseudoscalar decay constant $f_{PS}$ is given by\looseness-1
\begin{equation}
\label{defnfpi}
f_\pi = 
{Z_A \langle 0 | A_4^{\rm local} | \pi (\vec p )\rangle  \over {E_\pi(\vec p)}} \ ,
\end{equation}
where $Z_A$ is the renormalization constant connecting the 
lattice scheme to continuum \MSbar.  We study, in addition to the 2-point correlation
functions $\Gamma$, two kinds of ratios of correlators:
\begin{equation}
R_1(t) = {\Gamma_{SL}(t) \over { \Gamma_{SS}(t) }} \ ; \qquad
R_2(t) = {\Gamma_{SL}(t) \Gamma_{SL}(t) \over { \Gamma_{SS}(t) }} \,.
\end{equation}
Using either $\pi$ or $A_4$ for the smeared source $J$ gives 4 ways of
extracting $f_{PS}$. Two more ways are gotten by combining the mass
and amplitude of the 2-point correlation functions, $i.e.$ $\langle
A_4 P \rangle_{LS}$ and $\langle P P \rangle_{SS}$, and $\langle A_4
A_4 \rangle_{LS}$ and $\langle A_4 A_4 \rangle_{SS}$.

The data satisfy the following consistency checks: the six ways of
calculating $f_{PS}$ described above, and at each of the five values of
momentum, give results consistent to within $2\sigma$ \cite{DC95}.  (The one
exception is the $\vec p = (2,0,0)$ case where the signal is not good
enough to ascertain that we have fit to the lowest state.)  Even
though these estimates are correlated, consistent results do
indicate that fits have been made to the lowest state and reassure us
of the statistical quality of the data.  We use the $\vec p = (0,0,0)$ 
data in our final analysis as it has the best signal. 

\begin{figure} 
\figcaption{Plot of the Bernard-Golterman ratio $R$ for the quenched theory.}
{\epsfxsize=\hsize\epsfbox{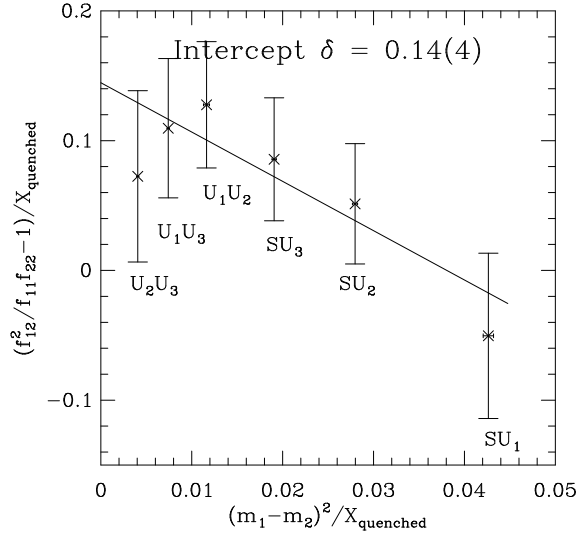}}
\vskip -10pt plus 10pt
\label{f_Rquen}
\end{figure}

\begin{figure} 
\figcaption{Plot of $R$ for the full theory.}
{\epsfxsize=\hsize\epsfbox{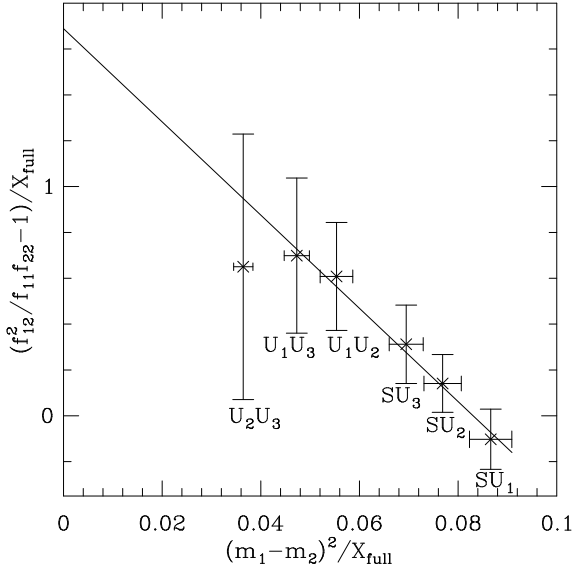}}
\vskip -24pt plus 10pt
\label{f_Rfull}
\end{figure}

\noindent{\bf Quenched approximation} When analyzed in terms of chiral 
perturbation theory (CPT), there are two consequences of using the
quenched approximation.  One, the coefficients in the quenched theory
are different from those in full QCD and uncalculable, and second,
Sharpe and collaborators \cite{SRScpt} and Bernard and Golterman
\cite{CbMgcpt} have pointed out that there exist extra chiral logs due
to the $\eta'$ as it is also a Goldstone boson in the quenched
approximation.  These make the chiral limit of quenched quantities
sick. To analyze the effects of $\eta'$ loops Bernard and Golterman
\cite{CbMgcpt} have constructed the ratio $R\equiv {f_{12}^2 /f_{11'}f_{22'}}$
%
applicable in a 4-flavor theory where $m_1 = m_{1'}$ and $m_2 =
m_{2'}$.  The advantages of this ratio in comparing full and quenched
theories is that it is free of ambiguities due to the cutoff $\Lambda$
in loop integrals and $O(p^4)$ terms in the chiral Lagrangian. CPT
predicts that 
\begin{eqnarray}
\label{eRquenched}
R^{Q} -1 &{}={}& \delta\ X_{quenched} + O((m_1-m_2)^2) \nonumber \\
R^{F} -1 &{}={}&         X_{full} + O((m_1-m_2)^2) \nonumber 
\end{eqnarray}
where $\delta \equiv {m_0^2 / 24 \pi^2 f_\pi^2}$ parameterizes the
effects of the $\eta'$, and $X_{quenched}$ and $X_{full}$ are given in 
\cite{chiralrg}.  At LATTICE94 the preferred fit (with 100 configurations 
and no $(m_1-m_2)^2$ term) was to the quenched expression which gave
$\delta=0.10(3)$ \cite{chiralrg}.  The need for including the
$(m_1-m_2)^2$ correction is shown in Figs.~\ref{f_Rquen} and
\ref{f_Rfull}. The fit to the quenched expression gives $\delta =
0.14(4)$, however, based on $\chi^2$, the fit to the full QCD
expression is preferred. The caveat is that the intercept is
$1.69(45)$ rather than unity.  Thus, we cannot resolve the effects of
$\eta'$ from normal higher order terms in the chiral expansion,
and neglect both in our analysis.

\begin{figure} 
\vskip -20pt plus 24pt
\figcaption{Plot of data for $f_\pi$ versus $m_{np}$.  The linear fit 
(solid line) is to the six \lilj\ points, with errors 
shown by the dotted lines. The dash-dot line is
a linear fit to the four $\ssli$ points. The vertical line at
$m_{np} \approx 0$ represents \mbar\ and the band at $m_{np} \approx
0.04$ denotes the range of $m_s$.}
{\epsfxsize=\hsize\epsfbox{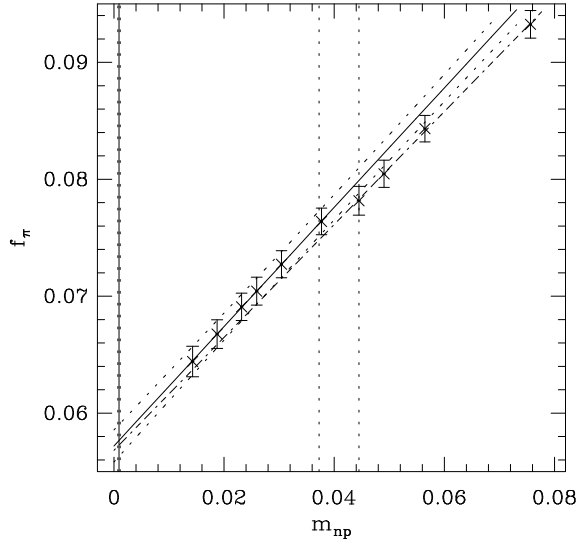}}
\vskip -24pt plus 10pt
\label{f_fpiextrap}
\end{figure}
\begin{figure} 
\figcaption{Extrapolation of heavy-light $f_{PS}$ to $\mbar$ 
for three cases, $C,\ S, \ U_1$, of ``heavy'' quarks. The linear fits
are to the three ``light'' $U_i$ quarks, and the fourth point
(light quark is $S$) is included to show the breakdown of the linear
approximation.}
{\epsfxsize=\hsize\epsfbox{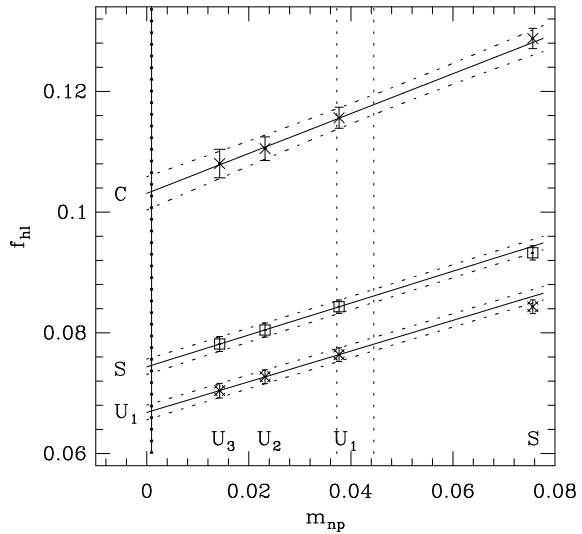}}
\vskip -24pt plus 10pt
\label{f_fpsvslight}
\end{figure}

{\bf Extrapolation to the physical quark masses}: The data, shown in
Fig.~\ref{f_fpiextrap}, indicates a break in the vicinity of $m_s$
between the non-degenerate \sli\ and degenerate $U_1U_1$ mesons at the
$1\sigma$ level, but no such break between the $U_1U_i$ and the
$U_2U_2$ cases.  We thus use \lilj\ points to extrapolate to $f_\pi$.
Note that even though the slopes for the two fits to \lilj\ and \ssli\
combinations are different, the values after extrapolation are
virtually indistinguishable.  In Fig.~\ref{f_fpsvslight} we show the
extrapolation for heavy-light mesons for three cases ($C,\ S, \ U_1$)
of ``heavy'' quarks. In all three cases we use a linear fit to the
three $U_i$ points for extrapolation to $\mbar$ as deviations from
linearity are apparent if the ``light'' quark mass is taken to be $S$
as shown by the fourth point at $m_{np} = 0.076$. To get $f_K$ we
interpolate to $m_s$ the result of the extrapolations of $SU_i$ and
$U_1U_i$ points to $\mbar$. For $f_D$ we extrapolate the three $CU_i$,
and for $f_{D_s}$ we simply interpolate between $CU_1$ and $CS$
points.

\noindent{\bf Results at \protect\boldmath $\beta=6.0$}: Our final results 
using $TAD1$ scheme along with estimates of statistical and various
systematic errors are given in Table~\ref{t_dcfinal}.  From the data
it is clear that systematic errors due to setting $m_c$, the lattice
scale, and $Z_A$ are now the dominant sources of errors.

\noindent{\bf Continuum Limit}: To extract results valid in the 
continuum limit we include data from the GF11
($\beta=5.7, 5.93, 6.17$) \cite{rGF11}, JLQCD ($\beta=6.1, 6.3$) \cite{rJLQCD},
and APE ($\beta=6.0, 6.2$) \cite{rAPE}\ Collaborations. We have
attempted to correct for as many systematic differences, however some, 
like differences in lattice volumes, range of quark masses analyzed,
and fitting techniques, remain.

Assuming that lattice spacing errors are $O(a)$, 
a linear fit versus $M_\rho a$ gives
\begin{eqnarray}
{f_\pi / M_\rho} \ &{}={}&\ 0.156(7) \qquad \hbox{\rm (expt. 0.170)}, \nonumber \\
{f_K / M_\rho}   \ &{}={}&\ 0.171(6) \qquad \hbox{\rm (expt. 0.208)}. \nonumber
\end{eqnarray}
with $\chi^2/_{dof} = 1.6$ and $1.7$ respectively. The change from the
GF11 results is marginal as the fit is still strongly influenced by
the point at $\beta=5.7$, which may lie outside the domain of validity
of the linear extrapolation.  A linear extrapolation excluding the $\beta=5.7$
data gives 
\begin{eqnarray}
{f_\pi / M_\rho}\ =\ 0.170(14) , \qquad {f_K / M_\rho}\ =\ 0.187(11)  , \nonumber 
\end{eqnarray}
with $\chi^2/_{dof} = 2.1$ and $1.9$ respectively.  Using
$m_s(M_\phi)$ would increase $f_K$ by $\sim 2\%$.  Given this
difference in the extrapolated value depending on whether the data at
$\beta=5.7$ is included or not makes it clear that more data are
required to make a reliable $a \to 0$ extrapolation.

\begin{table*} 
\caption{Our final results at $\beta=6.0$. 
All dimensionful numbers are given in $MeV$ with the scale set by
$M_\rho$. For the systematic errors due to $m_s,\ m_c,\ q^*$ we also
give the sign of the effect.  We cannot estimate the uncertainty due
to using the quenched approximation, or for entries marked with a ?.}
\vskip 6pt
{\hfil\input {t_dcfinal}\hfil}
\vskip -6pt
\label{t_dcfinal}
\end{table*}

\begin{figure} 
\figcaption{Extrapolation to the continuum limit of $f_D $ and 
$f_{D_s}$ (in MeV) data.  Our data is shown with the symbol octagon,
the plus points are from the JLQCD Collaboration \protect\cite{rJLQCD}, and the
diamonds label the APE collaboration \protect\cite{rAPE} data.}
{\epsfxsize=\hsize\epsfbox{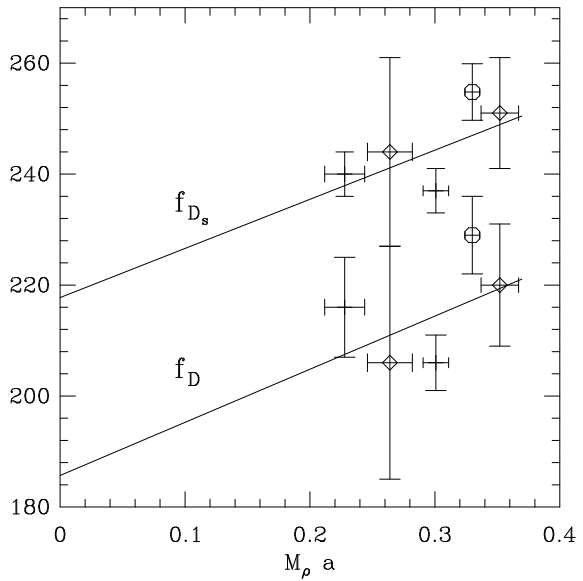}}
\vskip -24pt plus 10pt
\label{f_jlqcd}
\end{figure}

The $f_D$ and $f_{D_s}$ data at $\beta \ge 6.0$, in ${TAD1}$ scheme, and
using $m_s(M_K)$ are shown in Fig.~\ref{f_jlqcd}.  The APE
collaboration use $M_1$ for the meson mass. For consistency we have
shifted their data to $M_2$ using our estimates given in
Table~\ref{t_dcfinal}.  A linear extrapolation to $a = 0$ then gives
\begin{eqnarray}
f_D     &{}={}& 186(29) \MeV, \qquad f_{D_s} = 218(15) \MeV, \nonumber 
\end{eqnarray}
with $\chi^2/_{dof} = 2.2$ and $2.0$ respectively. Using $m_s(M_\phi)$
increases $f_{D_s}$ to $224(16)$ MeV. The quality of the fits are,
however, not very satisfactory. The bottom line is that in order to
improve the estimates the various systematic errors that have not been
included in the $a\to 0$ extrapolations presented above need to be
reduced.

\section{THE RARE DECAY $B\rightarrow K^*\gamma$.}

We discuss the applicability of heavy quark effective theory
(HQET) and pole dominance hypothesis (PDH) 
to extract the form-factors $T_1$ and
$T_2$ at $Q^2=0$ and $m_{heavy}=m_b$. The technical setup is the
same as described in \cite{Mfflat95} for the calculation of
semi-leptonic form-factors, and the quality of the signal is similar
to that for $D \to K^* l \nu$ decays.


\begin{figure} 
\figcaption{Three types of fits to test $Q^2$ behavior of $T_1$ 
using $CU_3 \to U_1U_3$ transition. HQET
suggests a dipole fit if one assumes PDH for $T_2$. The data prefer a 
pole fit but with the resonance mass smaller than the lattice measured 
value $M_{D^*}$ used to plot the data.}
\hbox{\epsfxsize=\hsize\epsfbox{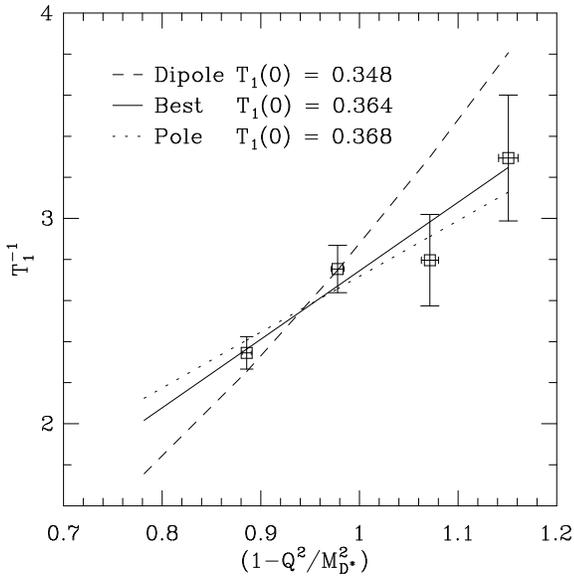}}
\vskip -24pt plus 10pt
\label{f_t1q2}
\end{figure}

\begin{figure} 
\figcaption{Fits to test $Q^2$ behavior of $T_2$.}
\hbox{\epsfxsize=\hsize\epsfbox{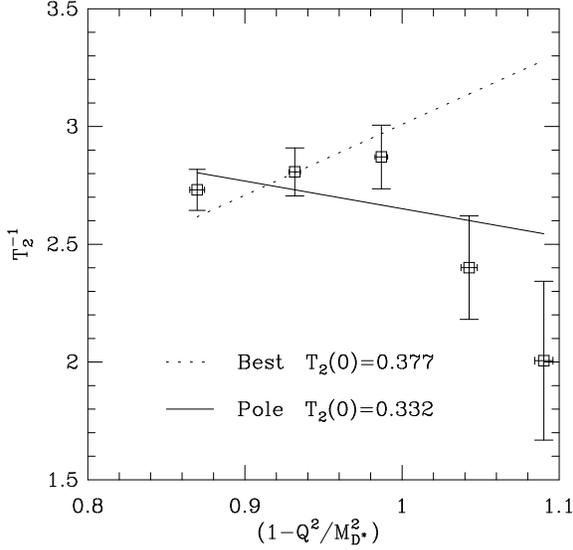}}
\vskip -12pt plus 10pt
\label{f_t2q2}
\end{figure}

\noindent{\bf PDH}: states that the $Q^2$ behavior of all form-factors is 
\begin{equation}
f(Q^2) \ = \ f(0) / (1 - Q^2/M^2)\,,
\label{e_poledominance}
\end{equation}
where $M$ is the mass of the nearest resonance with the right quantum
numbers.  To test PDH we make two kinds of fits: (i) single parameter
``pole'' fit where $M$ is the lattice measured value of the resonance
mass, (ii) two parameter ``best'' fit where $M$ and $f(0)$ are free
parameters. Typical examples of these fits are shown in Figs.~\ref{f_t1q2}
and \ref{f_t2q2}. Overall, $T_1$ is well described by the
``pole'' form, whereas $T_2$ has a ``flat'' $Q^2$ dependence.  We
take the ``best'' fit values for our final estimates.

\noindent{\bf HQET}: To leading order in $\alpha_s$ and in the mass of the heavy
quarks, HQET implies  (for heavy to heavy transitions) that the combinations 
\begin{equation}
{\sqrt{m_B m_{K^*}} \over m_B + m_K^*} T_1(Q^2) = 
{\sqrt{m_B m_{K^*}} \over m_B + m_K^*} {T_2(Q^2) \over 1 - {Q^2 \over 
(m_B + m_{K^*})^2}}
\label{e_hqet}
\end{equation}
are independent of the masses of the heavy quarks for fixed velocity
transfer. Since $T_1(Q^2=0) = T_2(Q^2=0)$ for all $m_q$, this HQET
relation and the PDH, Eq.~\ref{e_poledominance},
cannot hold simultaneously. In fact, for heavy quarks $(m_B + m_{K^*})
\approx m_{pole}$, therefore, if $T_2$ fits the pole form then $T_1$ must
be a `dipole'. Instead our data, as exemplified in Figs.~\ref{f_t1q2}
and \ref{f_t2q2}, prefer a flat $T_2$ and a pole behavior for $T_1$.

\begin{figure} 
\figcaption{%
Extrapolation of $T_1(Q^2=0)$ to $m_u$. The interpolation to $m_s$ for 
$T_1(B \rightarrow K^*\gamma)$ is done using the 
points labeled by squares ($s=S$) and octagons ($s=U_1$), while $T_1(B
\rightarrow \rho\gamma)$ is obtained by extrapolating the 
degenerate $q \bar q$ points (crosses).}
\hbox{\epsfxsize=\hsize\epsfbox{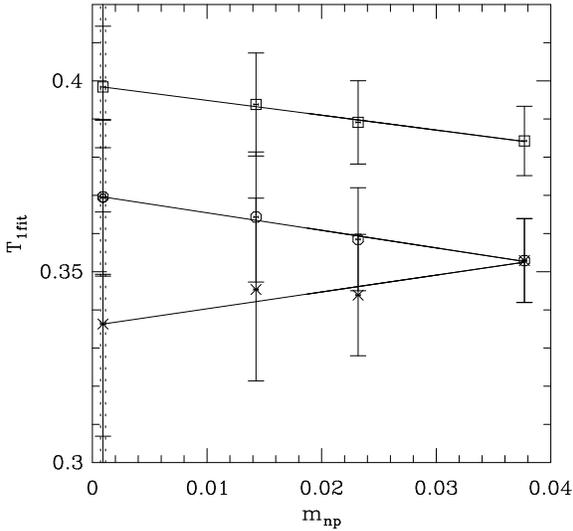}}
\vskip -24pt plus 10pt
\label{f_t1chiral}
\end{figure}

\begin{figure} 
\figcaption{%
Extrapolation of $T_2(Q^2=0)$ to $m_u$. The interpolation to $m_s$ for 
$T_2(B \rightarrow K^*\gamma)$ is done using the 
points labeled by squares ($s=S$) and octagons ($s=U_1$), while $T_2(B
\rightarrow \rho\gamma)$ is obtained by extrapolating the 
degenerate $q \bar q$ points (crosses).}
\hbox{\epsfxsize=\hsize\epsfbox{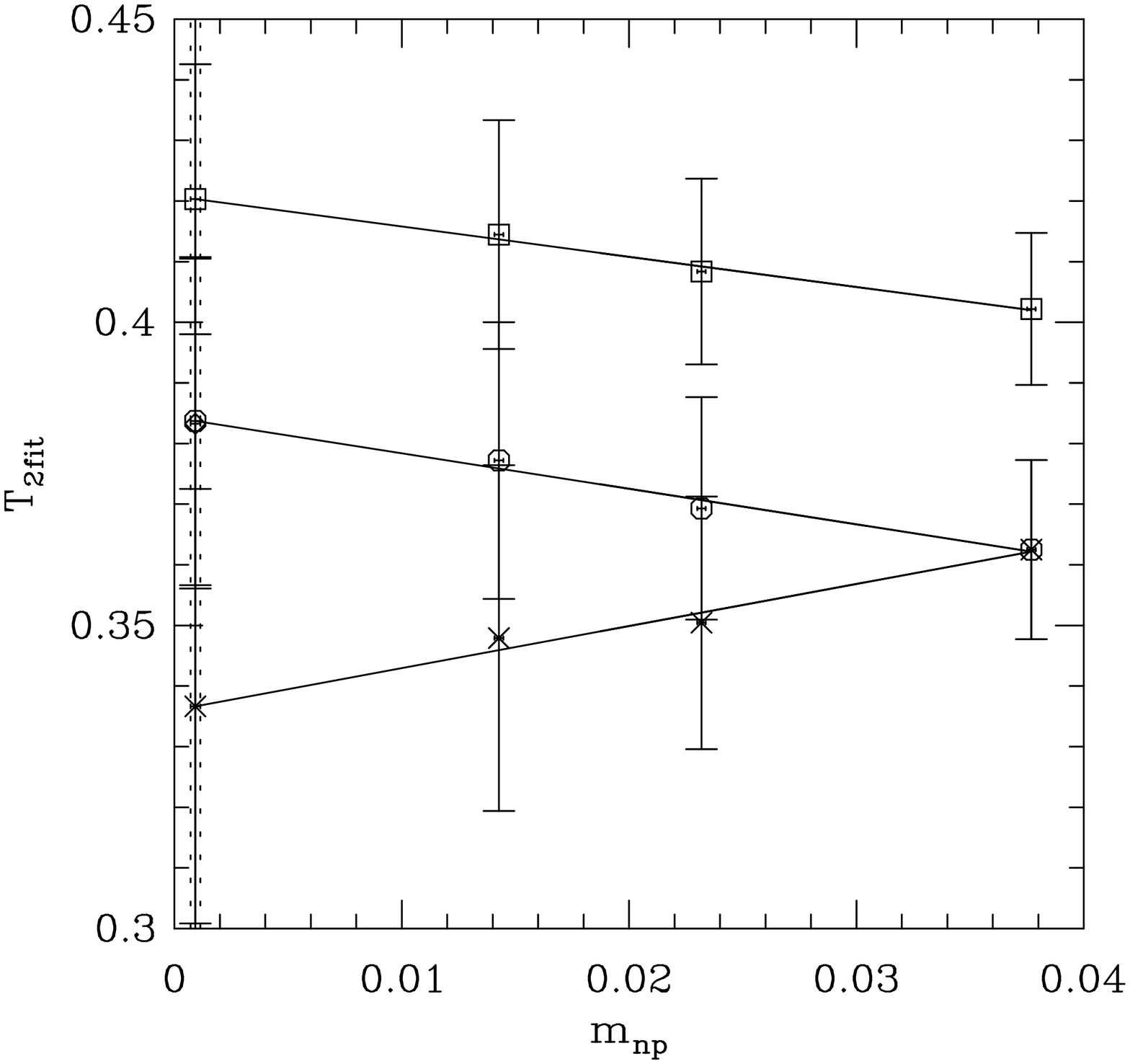}}
\vskip -24pt plus 10pt
\label{f_t2chiral}
\end{figure}

\noindent{\bf Dependence on quark mass}: 
Figures~\ref{f_t1chiral} and \ref{f_t2chiral} show examples of the
variation of $T_1(0)$ and $T_2(0)$ with quark masses. There is
significant dependence on the mass of the quark $C$ decays into (which
is a kinematic effect), and a slight dependence on $m_{spectator}$
resulting in the small increase in slope between $CU_i \to U_1U_i$ and 
$CU_i \to SU_i$ cases, which is consistent with HQET.

\begin{table} 
\caption{Estimates of form factors in 3 commonly used 
renormalization schemes defined in \protect\cite{DC95}. The data
satisfy $T_1(Q^2=0) = T_2(Q^2=0)$. The last four rows give 
$T(Q^2=0)$ extrapolated to $m_b$ using the 4 methods discussed in the
text.}
\vskip 6pt
\input{t_schemes}
\vskip -10pt plus 10pt
\label{t_schemes}
\end{table}

\noindent{\bf Extrapolation in \protect\boldmath $m_{heavy}$}: The 
need to extrapolate the results obtained at $m_{heavy} \approx m_c$ to
$m_b$ using HQET, Eq.~\ref{e_hqet}, introduces a very large
uncertainty as shown by the four ways of analyzing the data. Methods 1
and 2: we take the value of $T_2$ at zero recoil extrapolated to $m_s$
and $\mbar$ and scale it to $m_b$ using HQET. We can then estimate the value at
$Q^2=0$ assuming pole dominance holds for $T_2$ at $m_b$ (advocated by
A. Soni at this conference), or by using a ``flat'' behavior as shown
by data at $m_{heavy}=m_c$.  Method 3 (4): Scale $T_2(Q^2=0)$ ($T_1$)
assuming the joint validity of HQET and pole dominance. This implies
the scaling relations $T_2(Q^2=0) m_{heavy}^{3/2} = constant$ and 
$T_1(Q^2=0)m_{heavy}^{1/2} = constant$.  The results along with their
variation with the tadpole subtraction prescription, type of fit,
$m_s$, and the definition of heavy-light meson mass ($M_1$ or $M_2$)
are shown in tables \ref{t_schemes} and \ref{t_m1m2}.

\begin{table} 
\caption{Estimates in $TAD1$ scheme at $Q^2=0$. The variations give 
estimates of systematic errors.}
\vskip 6pt
\input{t_m1m2}
\vskip -18pt plus 10pt
\label{t_m1m2}
\end{table}

\noindent{\bf Results at \protect\boldmath$\beta=6.0$}: 
Methods 1,2 and 3,4 reflect the same contradiction. The value is
either $0.08-0.1$ or $0.23-0.25$ depending on what we assume for the
scaling behavior. With present data we 
assume that the flat $Q^2$ behavior for $T_2$ and pole dominance for
$T_1$ persists all the way upto the physical value of $m_b$. Then,
using the best fit, TAD1 subtraction prescription, $m_s(M_\phi)$, and
$M_1$ for the meson mass, we get $T_1=T_2=0.24(1)$. 
Further progress requires clarification of the $Q^2$
behaviour of the form factors and an estimate of the violations of
leading order HQET predictions.

\section{B-parameters}

We present an update on results for $B_K$, $B_7$, $B_8$ with Wilson
fermions evaluated in the NDR scheme with $TAD1$ subtraction
prescription. Note that both $q^*$ and the matching scale between the
lattice and continuum theories are taken to be $1/a$.
Thereafter, the results are run to $2\ \GeV$ using the 2-loop
relations, however the change is minimal. 

To analyze the lattice data (illustrated in Table~\ref{t_bkw}) we
consider the general form, ignoring chiral logs, of the chiral
expansion of the $\Delta S=2$ 4-fermion matrix elements with Wilson
fermions\looseness-1
\begin{eqnarray}
\left\langle \overline{K^0} \right| {\cal O}_{LL} \left| K^0 \right\rangle ={}&
\alpha + \beta m_K^2 + \gamma p_i p_f + \delta_1 m_K^4 \aftergroup\hfill \nonumber\\
{}+{}& \delta_2 m_K^2 p_i p_f + \delta_3 (p_i p_f )^2 + \ldots .
\label{eq:bkcpt}
\end{eqnarray}
This follows from Lorentz symmetry as $m^2$ and $p_i \cdot p_f$ are
the only invariants.

\begin{figure} 
\figcaption{Six parameter fit to the $B_K$ data.}
\hbox{\epsfxsize=\hsize\epsfbox{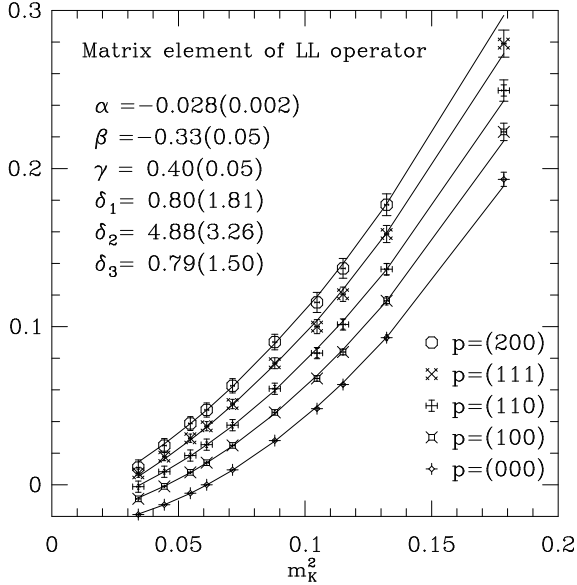}}
\vskip -24pt plus 10pt
\label{f_bkcpt}
\end{figure}

\noindent{\bf \protect\boldmath $B_K$}: The terms 
proportional to $\alpha, \beta$ and $\delta_1$ are pure lattice
artifacts due to mixing of the $\Delta S =2$ 4-fermion operator with
wrong chirality operators.  To isolate these terms we fit the data for
the lightest 10 mass combinations and for the 5 values of momentum
transfer using Eq.~\ref{eq:bkcpt} as shown in Fig.~\ref{f_bkcpt}.
(Similar values for the six coefficients are obtained from fits to the
6 lightest combinations.)  We find that the three
$\delta_i$ are not well determined; only $\delta_2$ is significantly
different from zero.  More important, the coefficients $\gamma, \delta_2,
\delta_3$ contain artifacts in addition to the desired physical pieces
which we cannot resolve by this method. We simply assume that the
1-loop improved operator does a sufficiently good job of removing
these residual artifacts. The result then is
\begin{eqnarray}
B_K (NDR, 1/a) = \gamma + (\delta_2 + \delta_3) M_K^2 = 0.65(10). \nonumber
\end{eqnarray}
A second way of extracting $B_K$ using Eq.~\ref{eq:bkcpt} is to combine pairs of points at
different momentum transfer:\looseness-1
\begin{eqnarray}
&{} \big( E_1 B_K(q_1) - E_2 B_K(q_2) \big) / (E_1-E_2)  = \nonumber \\
 &{} \qquad\qquad \qquad \qquad \gamma + \delta_2 m^2 + \delta_3 m (E_1+E_2).  \nonumber
\end{eqnarray}
This procedure directly removes $\alpha, \beta$ and $\delta_1$ but
requires a correction to the $\delta_3 m (E_1+E_2)$ piece, for which 
we use the value of $\delta_3$ extracted from the fit.  The results of
this analysis for the 10 light mass combinations
are given in the third column of Table~\ref{t_bkw}.  Interpolating to
$m_K$, we get $B_K(NDR, 1/a) = 0.66(9)$.


\noindent{\bf \protect\boldmath $\hat B_K$}: The 2-loop running of $B_K$ defines 
the renormalization group invariant quantity $\hat B_K$
\cite{guidoBK}
\begin{eqnarray}
\hat B_K &{}={}& \alpha_s(\mu)^{-2/\beta_0} \big(1 + \alpha_s(\mu) J / 4\pi \big) B_K(\mu)  \nonumber
\end{eqnarray}
where $J = (\beta_1 \gamma_0-\beta_0\gamma_1)/2 \beta_0^2 = 2.004$ for
$n_f=0$.  Under running, $B_K$ increases as $\mu$ is decreased. Thus
$B_K(NDR, 1/a) = 0.66(9)$ becomes $B_K(NDR, 2\GeV) = 0.67(9)$,
and $\hat B_K$ is $0.90(14)$. For comparison, the Staggered value
calculated at $\beta=6.0$ is $B_K(NDR, 2\GeV) = 0.67-0.71$ depending
on the lattice operators used \cite{sharpebk,jlqcdbk}.  An update on
staggered results and issues of extrapolation to $a\to 0$ have been
presented by JLQCD at this conference \cite{jlqcdbk}.

\noindent{\bf \protect\boldmath $B_D$}: 
The $CS$, $C U_i$ data show no significant variation with momentum transfer as 
shown in Table~\ref{t_bkw}.  The theoretical analysis of 
artifacts in heavy-light mesons has not yet been completed; indications
are that all 6 terms contribute.  Therefore we simply extrapolate the $CU_i$ data to
$\mbar$ to get $B_D(NDR, 1/a)=0.78(1)$ or $B_D(NDR, 2\GeV)=0.79(1)$.

\begin{table} 
\caption{$B_K$, $B_7$ and $B_8$ in NDR renormalization scheme at matching 
scale $\mu a=1$.\looseness=-1}
\vskip 6pt
\input {t_bkw}
\vskip -12pt plus 10pt
\label{t_bkw}
\end{table}

\noindent{\bf \protect\boldmath $B_7$ and $B_8$}: The chiral expansion is similar to 
Eq.~\ref{eq:bkcpt} with all 6 coefficients containing artifacts and
physical pieces.  Ignoring the artifacts we get $B_7(NDR,1/a) =
0.60(1)$ and $B_8(NDR,1/a) = 0.81(1)$ or with 2-loop running
$B_7(NDR,2\GeV) = 0.59(1)$ and $B_8(NDR,2\GeV) =0.81(1)$.  (We find
that $B_8$ is very insensitive to changes in $\mu$ as the running of
the matrix element is almost completely canceled by that of its
vacuum saturation approximation, while in the case of $B_7$ the two
add.)  Our estimate of $B_8$ is smaller than that used in the Standard
Model analysis of $\epsilon'/\epsilon$ \cite{guidoBK}. Since a smaller
$B_8$ means larger $\epsilon'/\epsilon$, the calculation of $B_8$ is
important phenomenologically. Work is in progress to understand and
remove various lattice artifacts and make our estimate more
reliable.\looseness-1

\end{document}

%% file: t_dcfinal.tex
\let\ifspace=\iffalse
\def\myskip{\omit&height1.5pt&%
\omit&&%
\omit&&%
\omit&&%
\omit&&%
\omit&&%
\omit&&%
\omit&&%
 &\cr}
\vbox{\hbox{\vbox{
\tabskip=0pt\offinterlineskip
\def\tlr{\noalign{\hrule}}
\halign {\strut#& \vrule\vrule#\tabskip=3pt&
  \hfil$#$\hfil&\vrule#&
  \hfil$#$\hfil&\vrule#&
  \hfil$#$\hfil&\vrule#&
  \hfil$#$\hfil&\vrule#&
  \hfil$#$\hfil&\vrule#&
  \hfil$#$\hfil&\vrule#&
  \hfil$#$\hfil&\vrule#&
  \hfil$#$\hfil&\vrule#\tabskip=0pt\cr\tlr
\omit&height1.5pt&\multispan{15   }&\cr
\myskip
&& 
&& {\rm Best }
&& {\rm Statistical\ \&}
&& {\rm Tuning }                
&& {\rm Tuning }                  
&&                    
&& {\rm Tuning}
&&                    
  &\cr
\myskip
&& 
&& {\rm Estimate }
&& {\rm Extrapolation}
&& m_s          
&& m_c
&& q^*
&& a\ (3\%)
&& Z_A                
  &\cr
\myskip\tlr
\omit&height0.5pt&\multispan{15   }&\cr\tlr
\myskip
&& f_\pi           
&& 134
&& 4
&& -
&& -
&& +2
&& 4
&& 10
  &\cr\ifspace\myskip&&
&& 
&& 
&& 
&&
&& 
&& 
&& 
&& 
  &\cr\fi\myskip\tlr
\myskip
&& f_K
&& 159 
&& 3
&& -3 
&& -
&& +3
&& 5
&& 10
  &\cr\ifspace\myskip&&
&& 
&& 
&& 
&&
&& 
&& 
&& 
&& 
  &\cr\fi\myskip\tlr
\myskip
&& f_D
&& 229
&& 7
&& -
&& +12
&& +4
&& 7
&& 14
  &\cr\ifspace\myskip&&
&& 
&& 
&& 
&& 
&& 
&& 
&& 
&& 
  &\cr\fi\myskip\tlr
\myskip
&& f_{D_s}
&& 260
&& 4 
&& -5
&& +15
&& +4
&& 8
&& 20
  &\cr\ifspace\myskip&&
&& 
&& 
&& 
&& 
&& 
&& 
&& 
&& 
  &\cr\fi\myskip\tlr
\myskip
&& f_K/f_\pi
&& 1.19
&& 0.02
&& -0.025
&& -
&& -
&& -
&& 0
  &\cr\ifspace\myskip&&
&& 
&& 
&& 
&& 
&& 
&& 
&& 
&& 
  &\cr\fi\myskip\tlr
\myskip
&& f_D/f_\pi
&& 1.71
&& 0.05
&& -
&& +0.09
&& -
&& -
&& ?
  &\cr\ifspace\myskip&&
&& 
&& 
&& 
&& 
&& 
&& 
&& 
&& 
  &\cr\fi\myskip\tlr
\myskip
&& f_{D_s}/f_D
&& 1.135
&& 0.021
&& -0.023
&& +0.006
&& -
&& -
&& 0
  &\cr\ifspace\myskip&&
&& 
&& 
&& 
&& 
&& 
&& 
&& 
&& 
  &\cr\fi\myskip\tlr
\cr}}}}

%% file: t_schemes.tex
\newcommand\ce[1]{\multicolumn{#1}{|c|}}
\begin{tabular}{|l|l|l|l|}
\hline
&TAD1&TAD$\pi$&TADU$_0$\cr\hline
$T_1         $&0.37(2) &0.39(2) &0.35(2) \cr
$T_2         $&0.39(3) &0.40(3) &0.36(2) \cr
$(1)    $&0.097(2)&0.100(2)&0.091(2)\cr
$(2)    $&0.234(8)&0.240(8)&0.218(8)\cr
$(3)    $&0.084(6)&0.086(6)&0.078(5)\cr
$(4)    $&0.236(12)&0.242(13)&0.220(11) \cr
\hline
\end{tabular}

%% file: t_m1m2.tex
\newcommand\ce[1]{\multicolumn{#1}{|c|}}
\setlength{\tabcolsep}{2.6pt}
\begin{tabular}{|l|l|l|l|l|l|l|}
\hline
&&\ce2{Pole}&\ce2{Best}\cr
\hline
&&$m_s(M_K)$&$m_s(M_\phi)$&$m_s(M_K)$&$m_s(M_\phi)$\cr
\hline
$T_1$&$m_1$&0.37(2)&0.38(2)&0.37(2)&0.37(2)\cr
     &$m_2$&0.37(2)&0.38(2)&0.37(2)&0.37(2)\cr
$T_2$&$m_1$&0.33(1)&0.34(1)&0.38(3)&0.39(3)\cr
     &$m_2$&0.33(1)&0.34(1)&0.38(3)&0.39(3)\cr
\hline
$(1) $&$m_1$&&&0.096(2)&0.097(2)\cr
              &$m_2$&&&0.100(2)&0.101(2)\cr
$(2) $&$m_1$&&&0.230(8)&0.234(8)\cr
              &$m_2$&&&0.239(9)&0.243(9)\cr
$(3) $&$m_1$&&&0.082(6)&0.084(6)\cr
              &$m_2$&&&0.095(6)&0.096(6)\cr
$(4) $&$m_1$&&&0.232(13) &0.236(12) \cr
              &$m_2$&&&0.242(13) &0.245(13) \cr
\hline
\end{tabular}

%% file: t_bkw.tex
\newcommand\0{hphantom{0}}
\newcommand\ce[1]{\multicolumn{#1}{c}}
\setlength{\tabcolsep}{1pt}
\begin{tabular}{lrrrrr}
\hline
&\ce3{$B_K$} &\ce1{$B_7$}&\ce1{$B_8$}\cr 
&\ce1{$(p=0)$}&\ce1{$(p=2)$}&\ce1{subtr.}&       &  \cr 
\hline
$CC    $&$ 0.92(1)$&$ 0.93(2)$&$         $&$ 0.86(1)$&$ 0.94(1)$\cr
$CS    $&$ 0.83(1)$&$ 0.85(2)$&$         $&$ 0.82(1)$&$ 0.94(1)$\cr
$CU_1  $&$ 0.81(1)$&$ 0.82(2)$&$         $&$ 0.81(1)$&$ 0.94(1)$\cr
$CU_2  $&$ 0.80(1)$&$ 0.80(3)$&$         $&$ 0.81(1)$&$ 0.94(1)$\cr
$CU_3  $&$ 0.79(1)$&$ 0.79(4)$&$         $&$ 0.80(1)$&$ 0.93(1)$\cr
$SS    $&$ 0.56(0)$&$ 0.64(2)$&$ 0.69(30)$&$ 0.72(1)$&$ 0.92(1)$\cr
$SU_1  $&$ 0.44(0)$&$ 0.57(2)$&$ 0.69(23)$&$ 0.70(0)$&$ 0.90(1)$\cr
$SU_2  $&$ 0.38(1)$&$ 0.53(2)$&$ 0.68(21)$&$ 0.68(0)$&$ 0.89(1)$\cr
$SU_3  $&$ 0.34(1)$&$ 0.51(3)$&$ 0.67(19)$&$ 0.67(1)$&$ 0.88(1)$\cr
$U_1U_1$&$ 0.24(0)$&$ 0.47(2)$&$ 0.68(16)$&$ 0.66(0)$&$ 0.87(1)$\cr
$U_1U_2$&$ 0.11(1)$&$ 0.41(3)$&$ 0.68(14)$&$ 0.64(0)$&$ 0.86(1)$\cr
$U_1U_3$&$ 0.00(1)$&$ 0.36(3)$&$ 0.66(12)$&$ 0.63(0)$&$ 0.84(1)$\cr
$U_2U_2$&$-0.09(1)$&$ 0.33(4)$&$ 0.67(11)$&$ 0.62(0)$&$ 0.84(1)$\cr
$U_2U_3$&$-0.29(1)$&$ 0.26(5)$&$ 0.65(10)$&$ 0.60(0)$&$ 0.82(1)$\cr
$U_3U_3$&$-0.60(2)$&$ 0.14(7)$&$ 0.63(09)$&$ 0.58(1)$&$ 0.80(1)$\cr
\hline
\end{tabular}